\documentclass[prl,superscriptaddress,showpacs,longbibliography,reprint,floatfix]{revtex4-1}

\usepackage{hyperref}
\usepackage{color}
\usepackage[usenames,dvipsnames]{xcolor}
\usepackage{amsmath}
\usepackage{amssymb}
\usepackage{graphicx}
\graphicspath{{graphics/}}
\usepackage{epsfig}
\usepackage{dcolumn}% Align table columns on decimal point
\usepackage{bm}% bold math
\usepackage{mathrsfs}
\usepackage{multirow}
\usepackage[all]{xy}
\usepackage{pbox}
\usepackage{verbatim}
\usepackage{braket}
\usepackage{dsfont}
\usepackage{hyperref}
\usepackage{color}
\usepackage[usenames,dvipsnames]{xcolor}
\usepackage{array}
\usepackage{makecell}
\usepackage{tabularx}
\newcolumntype{*}{>{\global\let\currentrowstyle\relax}}
\newcolumntype{^}{>{\currentrowstyle}}
\newcommand{\rowstyle}[1]{\gdef\currentrowstyle{#1}%
  #1\ignorespaces
}

\DeclareMathOperator{\Tr}{Tr}
\def\(({\left(}
\def\)){\right)}
\def\[[{\left[}
\def\]]{\right]}

\newcommand{\norm}[1]{\left\lVert#1\right\rVert}

\newcommand{\vb}{\mathbf{v}_{\rm b}}

\newcommand{\be}{\begin{equation}}
\newcommand{\ee}{\end{equation}}
\newcommand{\ben}{\begin{eqnarray}}
\newcommand{\een}{\end{eqnarray}}
\newcommand{\beq}{\begin{equation}}
\newcommand{\eeq}{\end{equation}}

\newcommand{\e}{{\text{e}}}

\renewcommand{\e}{\textnormal{e}}

\renewcommand{\Tr}{\textnormal{Tr}}

\usepackage{graphicx}
\graphicspath{{./Figures/}}

\usepackage{color,soul}

\begin{document}

\title{Machine learning time-local generators of open quantum dynamics}

\author{Paolo P. Mazza}
\affiliation{Institut f\"{u}r Theoretische Physik and Center for Quantum Science, Universit\"{a}t T\"{u}bingen, Auf der Morgenstelle 14, 72076 T\"{u}bingen, Germany}
\author{Dominik Zietlow}
\affiliation{Max Planck Institute for Intelligent Systems, Max-Planck-Ring 4, 72076 T\"ubingen, Germany}
\author{Federico Carollo}
\affiliation{Institut f\"{u}r Theoretische Physik and Center for Quantum Science, Universit\"{a}t T\"{u}bingen, Auf der Morgenstelle 14, 72076 T\"{u}bingen, Germany}
\author{Sabine Andergassen}
\affiliation{Institut f\"{u}r Theoretische Physik and Center for Quantum Science, Universit\"{a}t T\"{u}bingen, Auf der Morgenstelle 14, 72076 T\"{u}bingen, Germany}
\author{Georg Martius}
\affiliation{Max Planck Institute for Intelligent Systems, Max-Planck-Ring 4, 72076 T\"ubingen, Germany}
\author{Igor Lesanovsky}
\affiliation{Institut f\"{u}r Theoretische Physik and Center for Quantum Science, Universit\"{a}t T\"{u}bingen, Auf der Morgenstelle 14, 72076 T\"{u}bingen, Germany}
\affiliation{School of Physics and Astronomy, University of Nottingham, Nottingham, NG7 2RD, UK}
\affiliation{Centre for the Mathematics and Theoretical Physics of Quantum Non-Equilibrium Systems,
University of Nottingham, Nottingham, NG7 2RD, UK}

\begin{abstract}
In the study of closed many-body quantum systems one is often interested in the evolution of a subset of degrees of freedom. On many occasions it is possible to approach the problem by performing an appropriate decomposition into a bath and a system. In the simplest case the evolution of the reduced state of the system is governed by a quantum master equation with a time-independent, i.e.\ Markovian, generator. Such evolution is typically emerging under the assumption of a weak coupling between the system and an infinitely large bath. Here, we are interested in understanding to which extent a neural network function approximator can predict open quantum dynamics --- described by time-local generators  --- from an underlying unitary dynamics. We investigate this question using a class of spin models, which is inspired by recent experimental setups. We find that indeed time-local generators can be learned. In certain situations they are even time-independent and allow to extrapolate the dynamics to unseen times. This might be useful for situations in which experiments or numerical simulations do not allow to capture long-time dynamics and for exploring thermalization occurring in closed quantum systems.
\end{abstract}

\maketitle

\begin{figure}[t]
\centering
\includegraphics[scale=0.49]{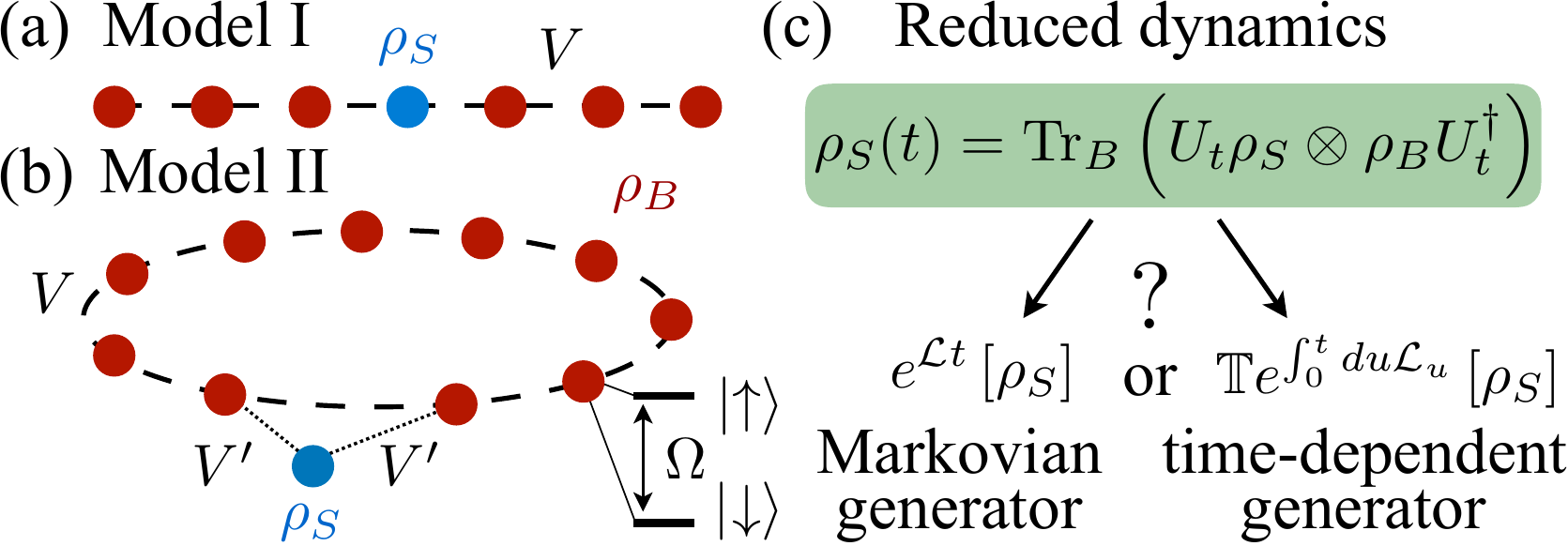}
\caption{{\bf Reduced local dynamics of a many-body quantum system.} The reduced state $\rho_S$ describes local degrees of freedom of a quantum spin system. The remainder of the system, $B$, takes the role of a ``bath'', with initial state $\rho_B$. We consider two scenarios: (a) Model I: spin chain with open boundary conditions, where the interaction strength $V$ and range $\alpha$ are varied [see Eq.~\eqref{eq:long-range}]; (b) Model II: spin chain with closed boundaries and nearest-neighbor interaction, where the system-bath coupling $V'$ and the initial temperature $\beta$ of the bath are modified. (c) The dynamics of the reduced state, $\rho_S(t)$, is obtained by tracing out the degrees of freedom of $B$ and to be learned by a neural network. The learned generator may be either Markovian (time-independent) or time-dependent.}
\label{Fig1}
\end{figure}

\emph{Introduction}---The investigation of isolated quantum systems out of equilibrium is a central problem in physics~\cite{ETHreview, Polkovnikov:2010yn}. Often, partial information that concerns a spatially localized subsystem ${S}$ is sufficient to study dynamical effects related to relaxation or thermalization. This information is encoded in the so-called reduced quantum state $\rho_{S}$, [c.f.~Fig.~\ref{Fig1}(a,b)], whose time dependence is obtained by ``integrating out" the remainder of the evolving many-body system, which is often referred to as bath ${B}$ [see Fig.~\ref{Fig1}(c)]. In certain settings, e.g.~for systems which are weakly interacting with a large environment \cite{breuer2002theory,gardiner2004quantum}, the dynamics of the reduced state is effectively described by a quantum master equation \cite{lindblad1976generators,gorini1976completely,alicki2007quantum}. In the simplest manifestation, the dynamics of the reduced quantum state $\rho_{S}$ is then governed by a time-independent generator $\mathcal{L}$ acting only on $S$:
\begin{eqnarray}
\rho_S(t):=\Tr_{ B}\left(U_t\left(\rho_S\otimes \rho_{ B}\right)U_t^\dagger\right)\approx e^{t\, \mathcal{L}}\left[\rho_S\right].
\end{eqnarray}
Here, $U_t$ is the unitary evolution operator of the full many-body system, the state $\rho_S\otimes \rho_B$ is the initial state of the many-body system in product form, and $\Tr_{B}$ indicates the average over the bath ${B}$. However, the generator does not need to be time-independent and, in most generic instances \cite{PhysRevLett.104.070406}, it may depend locally on time, $\mathcal{L}_t$, as sketched in Fig.~\ref{Fig1}(c). 

\begin{figure}[t]
    \centering
    \includegraphics[scale=0.5]{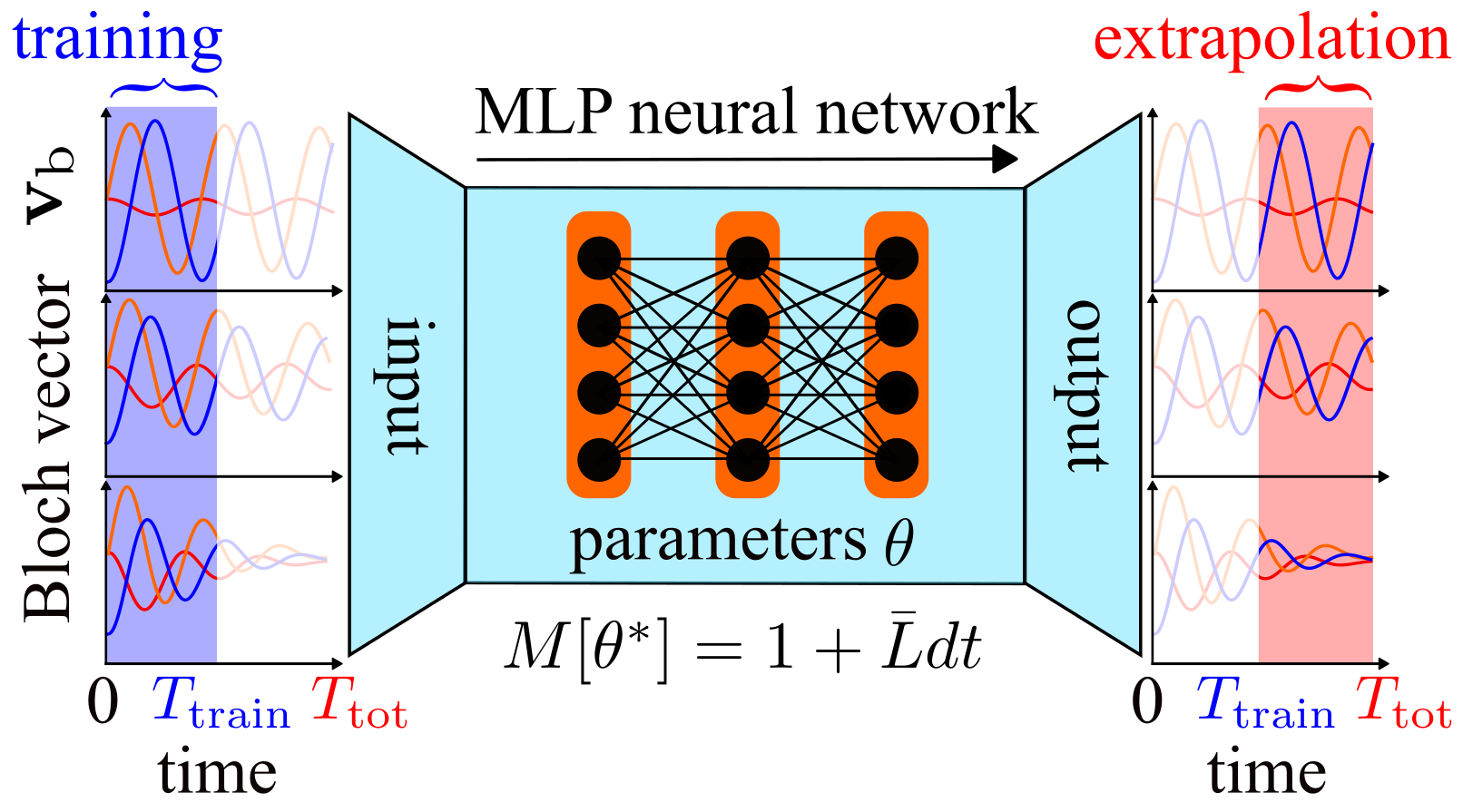}
    \caption{\textbf{Effective time-independent generator: Training and extrapolation}. To obtain a time-independent generator we use a multi-layer perceptron (MLP). The inputs are the time-dependent expectation values of the reduced system observables --- for a single spin contained in the Bloch vector $\mathbf{v}_{\rm b}$-- in a given training time window (shaded blue region) and for several initial Bloch vectors (top to bottom). The MLP learns the ``propagator" $M[\theta^*]$, which depends on the optimized network parameters $\theta^*$ but not on the actual time. Considering an Euler integration step of length $dt$, this provides a matrix $\bar{{L}}$ representing the action of an ``averaged" generator on the Bloch vector. When the reduced dynamics is Markovian, this learned generator allows for the network to make predictions for unseen future times (red shaded region).}
    \label{fig:Train}
\end{figure}

In this paper, we show how simple neural network architectures can learn time-local generators that govern the local dynamics of a closed quantum spin system. The input of the networks consists of the time-dependent average values of the reduced system observables from which the dynamical generator is estimated. Firstly, we consider an architecture which provides a time-averaged generator $\overline{\mathcal{L}}$ corresponding to an effectively Markovian description of the system dynamics. Our findings indicate that this can yield indeed a valid approximation for the generator even beyond typical scenarios justifying a Markovian weak-coupling assumption \cite{breuer2002theory,gardiner2004quantum}. Once the generator is known, the network can further be exploited to make predictions for times which go beyond the previously analyzed time frame, as is sketched in Fig.~\ref{fig:Train}. To study settings with a time-dependent generator, $\mathcal{L}_t$, we use a different neural network architecture which is based on so-called hypermodels and allows to assess the time-dependence of the generator of the reduced dynamics. 
Our work links to recent efforts aiming at understanding quantum dynamics with neural networks \cite{PhysRevB.99.214306,PhysRevLett.122.250502,PhysRevLett.122.250501,krastanov2020unboxing, ML1, ML2, CarleoSc, Carleo_Hartmann, liu2020, Luchnikov, Stefano,miles2020correlator}.  
Our approach, which uses machine learning tools and provides a directly interpretable object such as the physical generator of the dynamics, highlights a possible route for the application of neural networks in the study of the long-time dynamics of local observables in closed quantum systems. Moreover, it may find applications in the context of quantifying the degree of non-Markovianity in reduced quantum dynamics \cite{PhysRevLett.104.070406}.

\emph{Many-body spin models}---For illustrating our ideas, we consider two one-dimensional systems (in the following referred to as Model I and II) consisting of $N$ interacting spins, which are inspired by recent state-of-the-art quantum simulator experiments with Rydberg atoms~\cite{Bloch2012,kim2018,ebadi2020,Browaeys2020, Bloch_rev} or trapped ions \cite{islam2011,bohnet2016,joshi2020}. Model I has open boundary conditions [see Fig.~\ref{Fig1}(a)]. It features power-law interactions and is described by the Hamiltonian
\begin{equation}
    H_\mathrm{I} = \Omega\sum_{i=1}^N\sigma_i^x +  V\sum^N_{i<j}\frac{n_in_j}{|i-j|^\alpha}.
    \label{eq:long-range}
\end{equation}
Here, $n_i=(1+\sigma^z_i)/2$ is a projector on the ``up''-state of the $i$-th spin and $\sigma^{x,y,z}_i$ are Pauli matrices. The parameter $\alpha$ accounts for the interaction range, i.e.\ the power-law decay of the interaction potential, and $V$ controls the overall interaction strength with respect to the strength $\Omega$ of a transverse field term. The reduced system $S$, for this model, is the middle spin of the chain, as shown in Fig.~\ref{Fig1}(a). Model II is a closed spin chain [sketched in Fig.~\ref{Fig1}(b)], with nearest-neighbor interactions and Hamiltonian 
\begin{equation}
    H_\mathrm{II} =  \Omega\sum_{i=2}^N\sigma_i^x + V\sum_{j=2}^{N-1}n_jn_{j+1} + \Omega'\sigma_1^x + V^{'}n_1(n_2 +n_N).
    \label{eq:short_range}
\end{equation}
Here, the spin with label $1$ is singled out as reference spin (forming the system $S$) and the interaction strength with its neighbors is given by $V'$, while the interaction strength among all other spins is $V$. Moreover, we assume that the strength $\Omega'$ of the transverse field acting on the reference spin can be controlled. Both models capture a whole variety of different scenarios, including short- and long-range interactions, absence and presence of translational invariance as well as weak and strong coupling between local and bath degrees of freedom. They moreover encompass a number of standard scenarios often explored in quantum many-body physics, such as the so-called PXP-model \cite{sun2008,Ates2012,Papic2018short}, the Ising model in the presence of longitudinal and transverse fields, and all-to-all connected spin systems with resemblance to the Dicke model \cite{emary2003} or the Lipkin-Meshkov-Glick model \cite{dusuel2004}.

Our aim is to investigate the reduced dynamics of local degrees of freedom, as for instance a single spin $S$, as depicted in Fig.~\ref{Fig1}(a,b). To this end, we assume that the system is initialized at time $t=0$ in the state $\rho = \rho_S \otimes \rho_B$. In our studies concerning Model I, we assume the bath to be in the infinite temperature state, $\rho_B \propto \mathds{1}_B$. When considering Model II, instead, we assume a finite temperature situation with $\rho_B\propto e^{-\beta \tilde{H}_\mathrm{II}}$, where the Hamiltonian $\tilde{H}_\mathrm{II}$ is the one of Eq.~\eqref{eq:short_range} with $\Omega'=V'=0$, and $\beta$ is the inverse temperature. We let the state $\rho$ evolve according to the unitary dynamics $\rho(t) = U_t\rho U^\dag_t$, with $U_t = \e^{-iH t}$ from which we obtain the full dynamical information about the subsystem $S$. Throughout we will make use of the Bloch vector representation of the subsystem's density matrix: $\rho_S(t)=(\mathbf{v}_{\rm b}(t)\cdot \mathbf{\sigma})/2$ with $\mathbf{\sigma}=(\mathds{1},\sigma^x,\sigma^y,\sigma^z)$. 
Here the Bloch vector $\mathbf{v}_{\rm b}(t)=\left(1,\braket{\sigma_k^x}_t,\braket{\sigma_k^y}_t,\braket{\sigma_k^z}_t\right)$ is given in terms of the expectation $\braket{\cdot}_t$ with respect to the full quantum state at time $t$, and $k$ labels the system spin [see Fig. \ref{Fig1}(a,b)].

In the following, we will characterize the generator of this reduced dynamics; that is, we want to find a generator acting solely on $S$ that reproduces (or approximates) the dynamics of $\mathbf{v}_{\rm b}(t)$ for the two models under consideration [Fig.~\ref{Fig1}(a,b)]. For Model I we explore for which parameter combinations $(V,\alpha)$, the local generator, which is in principle time-dependent $\mathcal{L}_t$, can be approximated by an effective ``averaged'' time-independent one, $\mathcal{L}_t\approx \bar{\mathcal{L}}$. We remark here that, while $B$ acts as a fictitious bath for subsystem $S$, the situation described here is far from that of typical weak-coupling limits. Indeed, $B$ is a finite quantum system and its Hamiltonian has a discrete spectrum. As such, $B$ can hardly be thought of as an infinite Markovian bath of bosonic oscillators, whose state is unaffected by the interaction with $S$. The only aspect in common with a weak-coupling limit is that the interaction strength can be made small. However, this is not sufficient to argue that the local dynamics may be described by time-independent generators. 

For Model II we consider the case in which an ``isolated'' spin (on site $1$) interacts with $B$ with an independently tunable strength $V'$. Experimentally such a situation can be realized, e.g., by exploiting the distance dependence of dipolar interactions of atoms in Rydberg states \cite{Browaeys2020}. Also for Model II we investigate how well the generator of the reduced dynamics is approximated by a time-independent model when the initial state of $B$ (parametrized by the inverse temperature $\beta$) and the interaction strength $V^{'}$ is varied. In addition, for this scenario, we  use a hypermodel in order to analyze regimes that necessitate a time-dependent generator $\mathcal{L}_t$.

\emph{Network architectures and training}---In order to learn dynamical generators for the reduced dynamics of subsystem $S$, we employ machine learning algorithms. Our approach is completely data-driven, i.e.~the network has no prior information about the physical system. For learning time-independent generators, we use a linear multi-layer perceptron (MLP) architecture (see Fig. \ref{fig:Train}) which turns out to be the simplest possible one. Every data-point of our data-set $\mathcal D$ contains the triple $(\vb,\vb',t)$ where 
 $\vb := \vb(t)$ is the value of the Bloch vector at a given time-point $t$, and
 $\vb'$ is the value of the Bloch vector after a further discrete time step $dt$, i.e.\ $\vb':=\vb(t + dt)$. 
 
The input of our model is $\vb$, and the output is the vector $\mathbf{o}(t)=M[{\theta}]\vb$, where $M[{\theta}]$ is a $4\times4$-matrix (matching the dimension of the Bloch vector), that depends on the parameters ${\theta}$ of the network. To optimize the network, we introduce the following loss function, which is given by the norm of the difference between the next time step $\vb'$ and the output of the network:
\begin{equation}
C(\theta) = \mathbb E_{(\vb,\vb') \sim \mathcal D}\left[ \norm{M[{\theta}]\vb - \vb'} \right],
\end{equation}
where the expectation is taken over the training data $\mathcal D$.
Minimizing the above function thus provides a model $M[{\theta}^\ast]$ for the propagation of the Bloch vector over an infinitesimal time step $dt$. This model is related to the generator of the local dynamics: indeed, we have $M[{\theta}^\ast]= \mathds{1}+\overline{{L}}dt$, where $\bar{L}$ is a representation of the time-averaged generator $\overline{\mathcal{L}}$  %as 
acting on Bloch vectors. 
The training data-set is constructed by evolving the Bloch vector $\mathbf{v}_{\rm b}$ over a time interval $[0,T_\mathrm{train}]$. The trajectories are generated evolving $100$ randomly chosen initial states of the form
\begin{equation}
     \rho_0 = \frac{1}{2} \left(
    \begin{array}{@{}cc@{}}
    1+z & x - i y\\ 
    x + iy & 1-z
    \end{array}\right) \otimes \rho_B,
\end{equation}
with $x^2 + y^2 + z^2 <1$, i.e.~we consider non-pure initial states. We split our data-set such that we consider the $80\%$ of it for training, i.e.\ finding the best parameters of our model. The remaining $20\%$ form the validation set to assess the performance of the model built in the training phase. 

To encode and quantify the time-dependence of dynamical generators, $\mathcal{L}_t$, we use a different architecture and consider a so-called hypermodel, which computes network weights based on a context input. 
In our case, the context input is the time $t$ and a multi-layer perceptron (MLP) $F$ with non-linear activation functions transforms this input to the $4\times 4$ matrix of the generator $M[ F(t;\theta)]$. 

The data-set is organized in the same way as for the time-independent model. The only difference with respect to the previous case is that we also pass the information about the actual time step $t$. The network parameters $\theta$ are optimized by minimizing the loss function 
\begin{equation}
    C_\mathrm{H}(\theta) = \mathbb E_{(\vb,\vb',t) \sim \mathcal D} \left[ \norm{M[F(t;\theta)]\vb - \vb'}\right]. 
\end{equation}
By training the model in this way we obtain a time-dependent representation of the propagator of the local dynamics, i.e.\ $M[F(t;\theta^*)]= \mathds{1} + L_tdt$. For the training set, we choose quantum states as in the previous case. For the evaluation data-set, instead, we consider
\begin{equation}
\rho_0=  
\left(
\begin{array}{@{}cc@{}}
1-c & 0\\ 0&  c
\end{array}\right)\otimes \rho_B,
\label{eq:initial_condition}
\end{equation}
with $c$ being a random number, $c\in(0.01,0.7)$\footnote{See supplemental material for details}.

\begin{figure}[t]
    \centering
    \includegraphics[width=\columnwidth]{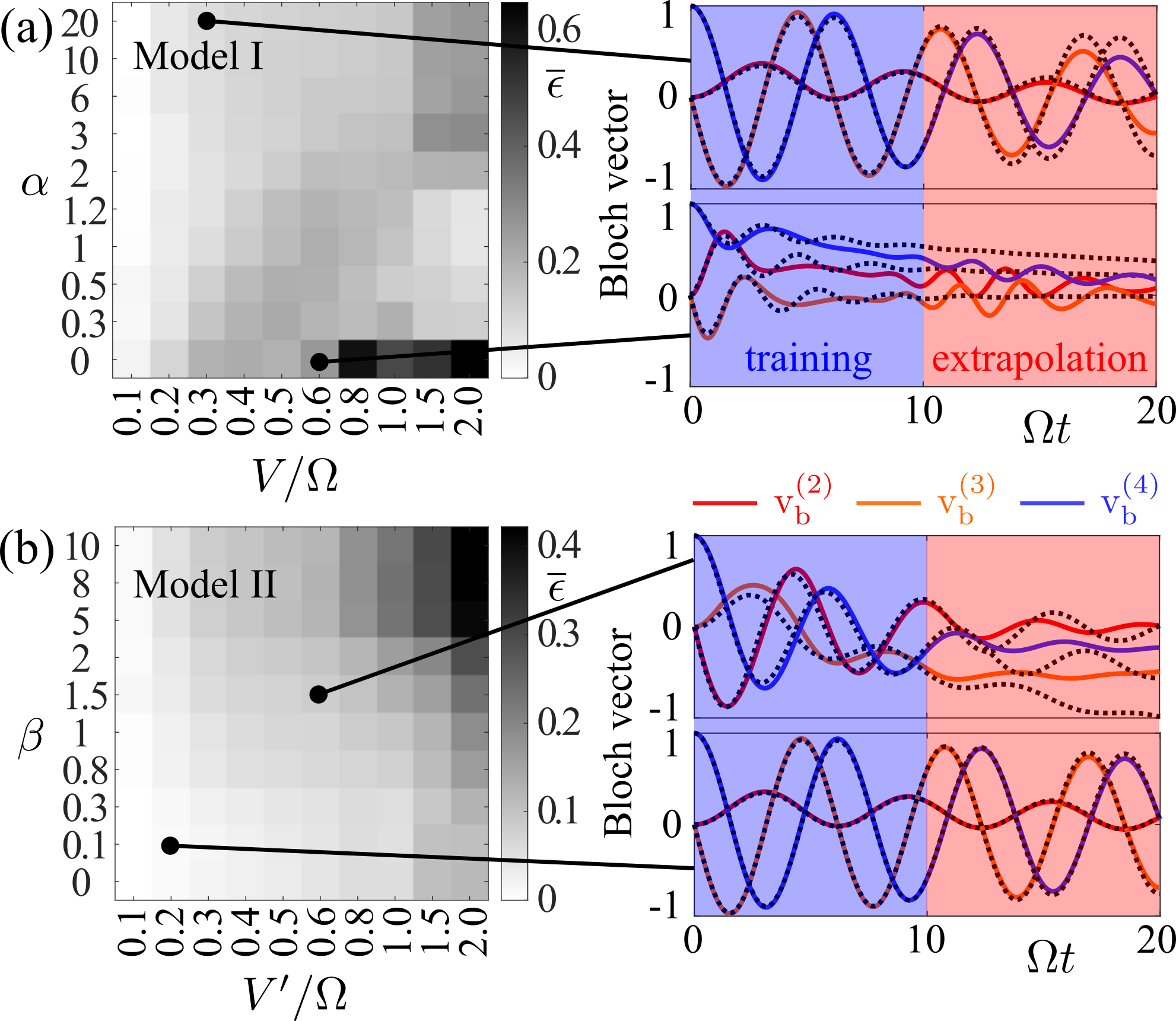}
    \caption{\textbf{Time-independent generator.} (a) Model I:
    Comparison between the exact dynamics of the Bloch vector and the one obtained by learning an effective time-independent generator with neural networks. The density plot shows the time-averaged norm difference, $\overline{\epsilon}$ [Eq. (\ref{eq:Error}) averaged over $5$ randomly selected values $c$ of the initial conditions, see Eq. (\ref{eq:initial_condition})], for various parameter choices $(V,\alpha)$. The data is calculated for $N=9$ spins with a training time of $T_\mathrm{train}=10/\Omega$, $dt = 0.01/ \Omega$ and a total time $T_\mathrm{tot}=20/\Omega$ (see main text for details). We also show two examples for the Bloch vector evolution, where we choose $c=1$ and $T_\mathrm{tot}=20/\Omega$ for illustration. Here solid curves correspond to the components of the Bloch vector propagated with the learned time-averaged generator. The dashed lines correspond to the numerically exact solution. (b) Same as panel (a), but for Model II and the parameter set $(V',\beta)$.}
    \label{fig:models}
\end{figure}
\emph{Time-independent generators}---With the above numerical approach, we can now investigate under which circumstances the generator of the dynamics of the subsystem $S$ is time-independent. We consider first the open spin chain [see Fig.~\ref{Fig1}(a) and Eq.~\eqref{eq:long-range}] and explore different values of $V$ and $\alpha$. To test the performance of the model after the training, we study the \emph{time-averaged} norm difference between the exact Bloch vector $\mathbf{v}^{\rm ex}_{\rm b}$ --- obtained by simulating the full many-body spin chain dynamics --- and the time-evolution of the same quantity as predicted by our model $\mathbf{v}^{\rm mod}_{\rm b}$:
\begin{equation}
     \epsilon=\frac{1}{T_{\rm tot}}\int_0^{T_{\rm tot}}\!\!dt\norm{\mathbf{v}^{\rm mod}_{\rm b}(t)-\mathbf{v}_{\rm b}^{\rm ex}(t)}\, ,
     \label{eq:Error}
 \end{equation}
computed by numerical integration. 

As shown in Fig.~\ref{fig:models}(a), the error $\epsilon$ is small for short-range (large $\alpha$) and weakly interacting (with small ratio $V/\Omega$) spin chains. This means that our time-independent model captures the relevant features of the reduced dynamics, and suggests that for finite systems with sufficiently weak interactions, the dynamics of local degrees of freedom can indeed be effectively described by a \emph{time-independent} generator $\bar{\mathcal{L}}$. Furthermore, for this Markovian regime of the dynamics, our model allows to make predictions for times exceeding the training time.

We now turn to the discussion of Model II [see Fig.~\ref{Fig1}(b) and Eq.~\eqref{eq:short_range}]. Using the same procedure explained above, we can learn the time-averaged dynamical generator for subsystem $S$. In Fig.~\ref{fig:models}(b) we summarize the results of this analysis. We display the time-averaged error, as defined in Eq.~\eqref{eq:Error}, for different parameter choices, which in this case are the (inverse) temperature $\beta$ of the initial state $\rho_B$ and the interaction strength $V^{'}$. We furthermore take $\Omega'=\Omega$. As shown in the figure the dynamics is  well approximated by a time-independent generator, $\bar{\mathcal{L}}$, for weak interactions $V'$ and for small $\beta$. Here also the extrapolation to times that exceed $T_\mathrm{train}$ is possible. The data in Fig.~\ref{fig:models} shows that in certain parameter regimes the approximation of the dynamics through a time-independent generator does not work well. Here significant deviations are even observable during the training period.

\begin{figure}[t]
    \centering
    \includegraphics[width=\columnwidth]{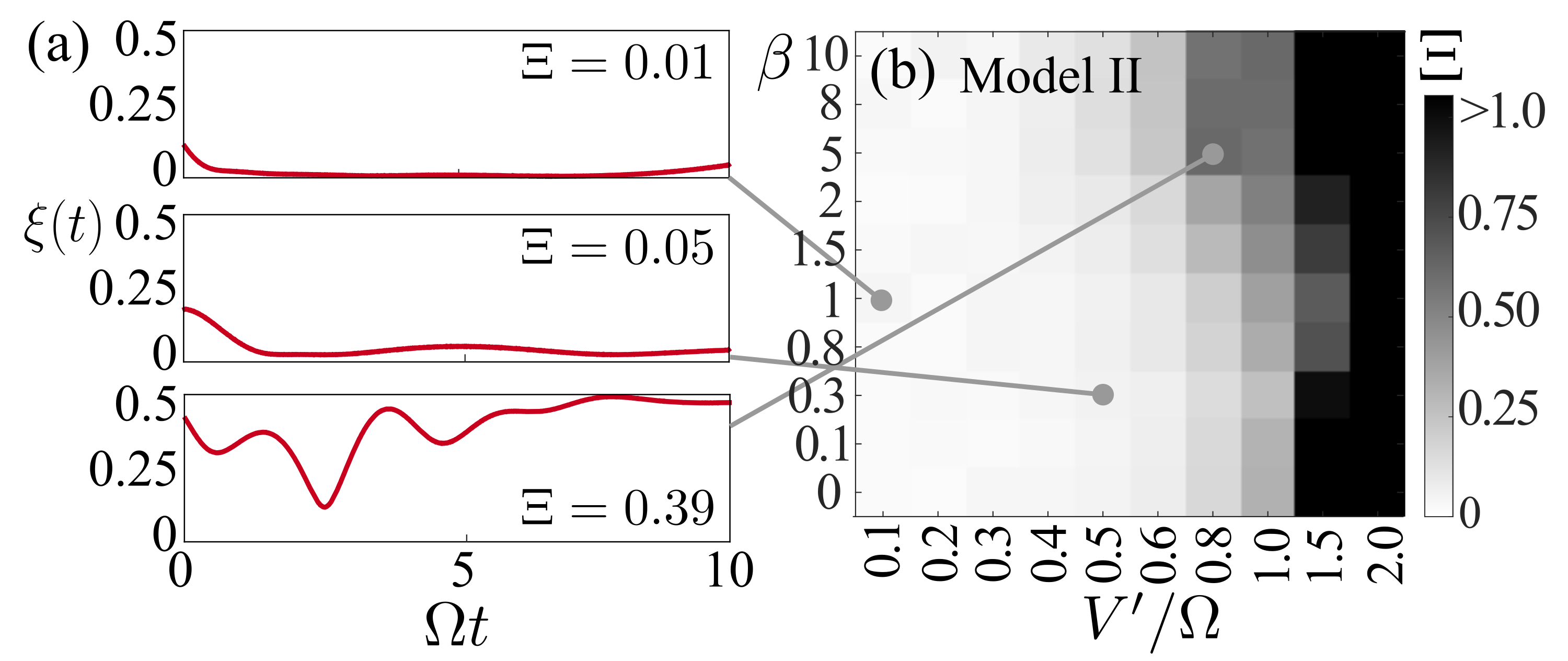}
    \caption{\textbf{Hypermodel and time-dependent generator.}  Hypermodels allow to encode time-dependent generators $\mathcal{L}_t$. The time-dependence is quantified through the derivative of $L_t$ with respect to time, expressed by the positive quantity $\xi(t)$ (see text for definition). Panel (a) shows three example curves corresponding to Model II ($N=9$). The function $\xi(t)$ depends only on the parameters of the model and is independent from the initial condition chosen. Generally, the stronger the interaction $V'$, the more pronounced the time dependence. This trend is clearly visible in panel (b), where we show the averaged value of $\xi(t)$ in the interval $0$ to $T=10/\Omega$, denoted as $\Xi$. In order to obtain the density plot we averaged, also in this case, over $5$ different values of $c$ in Eq.~\eqref{eq:initial_condition}, also in this case $dt = 0.01 \Omega$.  }
    \label{fig:hypermodel}
\end{figure}
\emph{Time-dependent time-local generators}---We are now interested in analyzing how strongly the dynamical generator depends on time, in those instances where the model introduced in the previous section fails. To this end, we adopt a hypermodel which can take as input the information about the running time. In this way, the neural network is able to learn an optimal parametrization of the ``propagator" $M[F(t;\theta^*)]= \mathds{1}+{{L}_t}dt$ for the Bloch vector dynamics which explicitly depends on time. Here, the matrix $L_t$ encodes the action of the generator on the Bloch vector and, essentially, implements the differential equations obeyed by the entries of the Bloch vector $\mathbf{v}_{\rm b}$. This architecture allows to learn and accurately reproduce the dynamics of Model II for all studied parameter regimes (see Supplemental Material for examples). This is not surprising, but gives us a handle for analyzing the time-dependence of the dynamical generator $L_t$. To this end we consider the positive quantity
$\xi(t)=\sqrt{{\rm Tr}(\dot{L}_t^\dagger \dot{L}_t)}$, where the dot denotes the derivative with respect to time. This can only be zero when $\dot{L}_t=0$, i.e.~when the action of the generator is not depending on the running time. To quantify an overall time-dependence within a time window $[0,T]$, we define the time-averaged value $\Xi=\frac{1}{T}\int_0^Tdt \, \xi(t)$. In Fig. \ref{fig:hypermodel} we show the corresponding data. For strong interactions $V'$, $\xi(t)$ is non-zero throughout which indicates a strong explicit time-dependence. This is also reflected in a large $\Xi$. For weak interactions, on the other hand, $\xi(t)$ remains small for all times considered. The oscillation we attribute to the finite size of the system. This confirms that, in such parameter regime, it is indeed possible to approximate the dynamics of the Bloch vector by means of a time-independent matrix, as considered in the previous section.

\emph{Conclusions}---We have presented two simple examples of neural network architectures that can learn the dynamical features of reduced quantum states. When such time evolution is effectively Markovian, the network can find a suitable approximation for the generator of the local dynamics. Here one can extrapolate the dynamics of reduced degrees of freedom to times that were unexplored during the training procedure. This possibility is particularly promising for applications in combination with tensor networks, which perform extremely well for short times. A neural network could learn the time-independent generator during this time interval and then possibly extrapolate to long times. In principle, this might enable the exploration of the onset of stationary or thermalization regimes. When the dynamics is not of Markovian type, we have shown that hypermodels can recover the generator of the reduced quantum dynamics. This allows to quantify non-Markovian effects \cite{BENATTI2016381}, which manifest in an explicit time-dependence of the dynamical generator Ref.~\cite{PhysRevLett.104.070406}. Furthermore, being able to reconstruct the generator of the reduced dynamics, as we have done here with hypermodels, makes it possible to explore different measures of non-Markovianity based on non-divisibility criteria for quantum dynamical time-evolutions \cite{PhysRevLett.103.210401,PhysRevA.81.062115,Breuer_2012,Rivas_2014}.

A possible future development in this regard could be the application of more advanced machine learning methods for learning time correlations in time series. This can be achieved with models borrowed from language studies (e.g. long-short term memory  recurrent neural networks (LSTM) \cite{LSTM} and transformers \cite{Transformers}) or with algorithms geared towards more interpretable models by either learning analytical expressions of the differential equation \cite{SahooLampertMartius2018:EQLDiv} or by modelling the time-evolution dynamics directly \cite{NODE}.

\acknowledgements
\emph{Acknowledgments}---{We acknowledge financial support from the Deutsche Forschungsgemeinschaft (DFG, German Research Foundation) under Germany’s Excellence Strategy – EXC-Number 2064/1 – Project number 390727645. IL acknowledges funding from the “Wissenschaftler Rückkehrprogramm GSO/CZS” of the Carl-Zeiss-Stiftung and the German Scholars Organization e.V., and through the DFG projects number 428276754 (SPP GiRyd) and 435696605. The authors thank the International Max Planck Research School for Intelligent Systems (IMPRS-IS) for supporting DZ.
}
% Supplementary
\bibliographystyle{apsrev4-1}
\bibliography{bib}

\begin{widetext}

\newpage 

\begin{center}
\textbf{\large Supplemental material: Machine learning time-local generators of open quantum dynamics} \\ \label{Sec:Supp}
\end{center}
\setcounter{equation}{0}
\setcounter{figure}{0}
\setcounter{table}{0}
\setcounter{page}{1}
\makeatletter
In this supplemental material we report a more detailed analysis concerning the effectiveness of the hypermodel in predicting the dynamics of the Bloch vector within the training time-window. In addition, we provide the hyper-parameters of the used models.

\section{Hypermodel vs. exact dynamics}
In the main text we analyze the time-dependence of time-local generators parametrizing the ``propagator'' $M[F(t; \theta^\ast)]=\mathds{1}+dtL_t$ for the Bloch vector, using a hypermodel architecture. The matrix $L_t$ encodes all the information concerning the dynamics of local (one-site) observables. As stated in the main text, a suitable measure to quantify the dependence on time of this matrix is given by $\Xi=\frac{1}{T}\int_0^Tdt \, \xi(t)$, where $\xi(t)=\sqrt{{\rm Tr}(\dot{L}_t^\dagger \dot{L}_t)}$. However, this can be a faithful measure of the time-dependence of the physical generator,  only if $L_t$ correctly implements the dynamics of the Bloch vector. It is thus important to check that the prediction of the hypermodel matches the exact dynamics, in the time window $[0,T]$ in which we want to analyze the generator  $L_t$. To this end, we report in Fig.~\ref{fig:comparison_hyp} a detailed comparison between the predicted time evolution of the Bloch vector components and the exact ones for six different parameter regimes.
\begin{figure}[h!]
    \centering
    \includegraphics[scale=0.4]{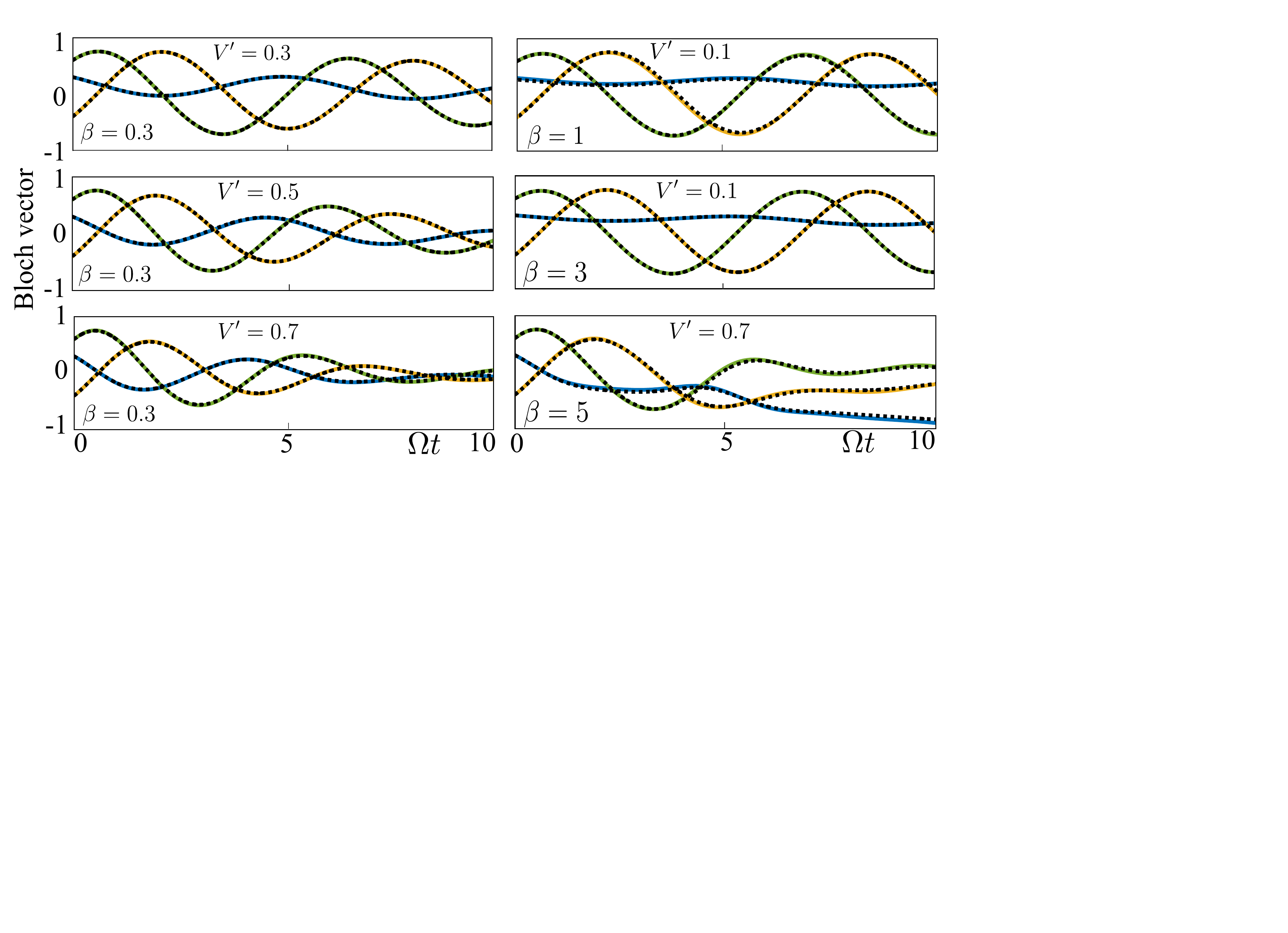}
    \caption{\textbf{Hypermodel accuracy}: Comparison between the exact dynamics and the dynamics predicted by the hypermodel for initial state $\rho_0$ reported in Eq.~\eqref{eq:sup_init} and for $N=7$. As it can be observed the agreement between the model prediction (dotted line) and the exact dynamics (full line) is excellent. This fact enables us to use the local operator learned using the hypermodel for studying the time-dependence of $L_t$. }
    \label{fig:comparison_hyp}
\end{figure}
We start from a generic initial state of the form
\begin{equation}
    \rho_0 = \frac{1}{2} \left(
    \begin{array}{@{}cc@{}}
    1+z & x - i y\\ 
    x + iy & 1-z
    \end{array}\right) \otimes \rho_B,
    \label{eq:sup_init}
\end{equation}
with $x^2 + y^2 + z^2 <1$, in particular we choose $x=0.6$, $y=0.3$ and $z=0.4$, and we consider 
different interactions and initial bath temperatures. In the training time-window $[0,T]$, we observe an excellent agreement between the two curves for $T=10/\Omega$. In Fig.~\ref{fig:comparison_hyp_der} we report the time evolution of the quantity $\xi(t)$ for the same parameter regimes.  %analyzed previously.
\begin{figure}[h!]
    \centering
    \includegraphics[scale=0.4]{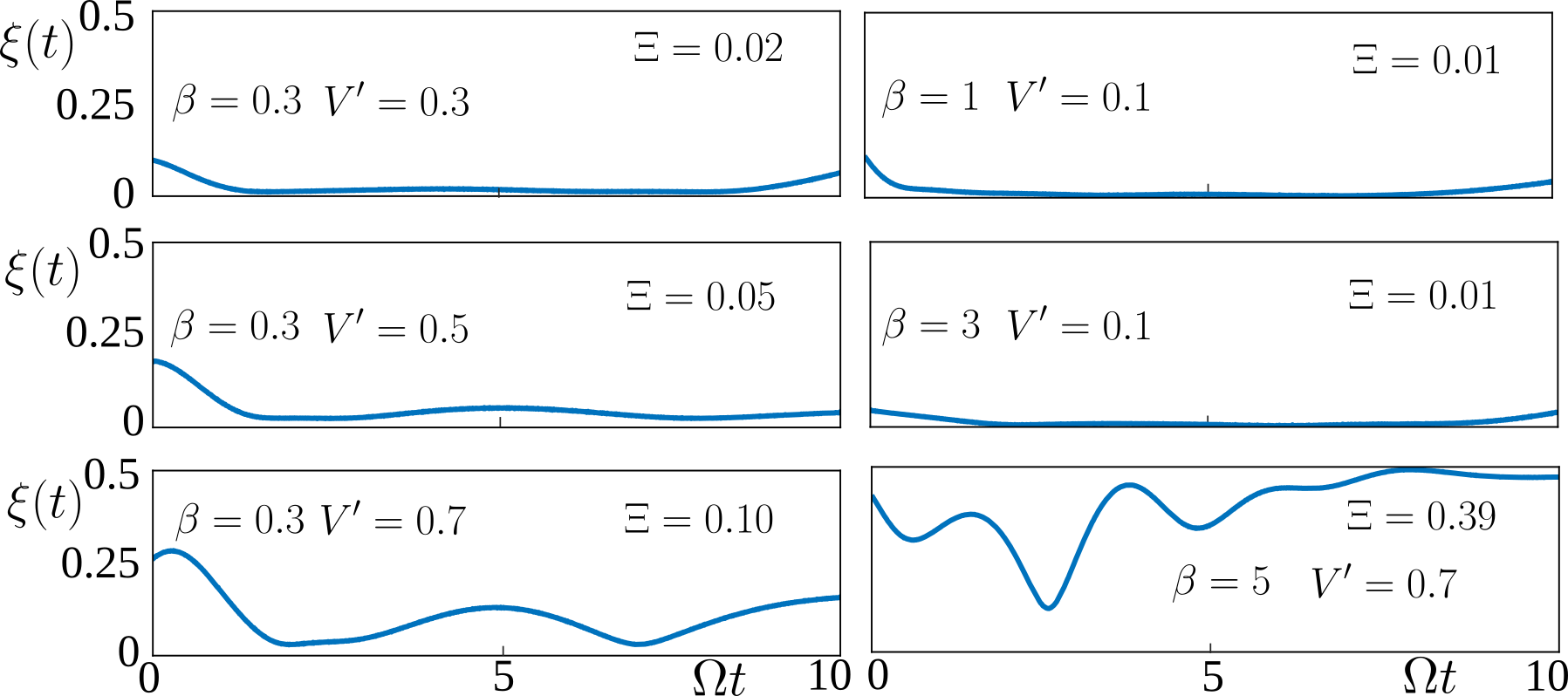}
    \caption{\textbf{Time-dependence}: Behavior of the quantity $\xi(t)$ for the same parameter regimes of Fig.~\ref{fig:comparison_hyp}. The quantity $\Xi$ grows with the interaction, meaning that the generator of the local dynamics $L_t$ becomes more time-dependent.}
    \label{fig:comparison_hyp_der}
\end{figure}
As it can be observed for weakly interacting systems the generator $L_t$ is almost time-independent. In this regime, as discussed in the main text, it is possible to predict the local dynamics using simple time-independent architectures. Increasing the interaction, the parameter $\Xi$ grows accordingly signaling that the dependence on time of $L_t$ becomes stronger.\\
The training procedure for the hyper-model is slightly different with respect what we have done for the time-independent case, in particular it is different the differentiation between training and evaluation set. In this case, in fact, we generate two different types of data-sets, the first data-set for training consists in $100$ trajectories with randomly chosen initial state $\rho_0$ of the the same form reported in Eq.~\eqref{eq:sup_init}. For evaluation we select, instead, randomly  data-points from trajectories generated from initial states $\rho_0^\mathrm{ev}$ of type
\begin{equation}
     \rho_0^\mathrm{ev} =  
     \left(
    \begin{array}{@{}cc@{}}
    1-c & 0\\ 0&  c
    \end{array}\right)\otimes \rho_B,
\end{equation}
with $0.01<c<0.7$. In this way we are performing a separation of the data-sets in the Hilbert space. This is contrary to what we had for the time-independent case where the data-set was given by separated time-intervals of the same trajectory.
\section{Network architectures}
\begin{table}[h!]
    \centering
    \begin{tabularx}{\textwidth}{*p{12em}|^p{13em}|^p{11em}|^p{16em}}\rowstyle{\bfseries}
        Setting & Architecture & Optimizer & Dataset \\\hline\hline
        \textbf{Time-independent} & 
        \makecell[tl]{Linear perceptron\\(no hidden layers)} & 
        \makecell[tl]{Adam \\
        Learning rate: $10^{-3}$\\
        Betas: $0.9$, $0.999$\\
        Epsilon: $10^{-8}$\\ 
        Batch size: $256$ \\
        Batches per epoch: $512$\\
        Epochs: $5$} &
        \makecell[tl]{Trajectories (train / val): $100$ / $20$ \\
            Total time ($\Omega\, T$): $10$\\
            Time step ($\Omega\, dt$): $0.01$
        }\\\hline
        \makecell[tl]{\textbf{Time-dependent}\\(hyper-model)} & 
        \makecell[tl]{Hidden layers: $3$\\
        Nonlinearity: \textit{tanh}\\
        Output nonlinearity: \textit{none}} & 
        \makecell[tl]{Adam \\
        Learning rate: $10^{-3}$\\
        Betas: $0.9$, $0.999$\\
        Epsilon: $10^{-8}$ \\      
        Batch size: $256$ \\
        Batches per epoch: $256$\\
        Epochs: $500$} &
        \makecell[tl]{Trajectories (train / val): $100$ / $20$ \\
            Total time ($\Omega\, T$): $10$\\
            Time step ($\Omega\, dt$): $0.01$
        }
        
    \end{tabularx}
    \caption{Overview over the architecture, optimizers, training- and system-parameters for Model I and Model II (top) and the hyper-model (bottom)}
    \label{tab:my_label}
\end{table}

\end{widetext}

\end{document}